# Travelling-wave single-photon detectors integrated with diamond photonic circuits - Operation at visible and telecom wavelengths with a timing jitter down to 23 ps

Patrik Rath[a,*], Andreas Vetter[a,b], Vadim Kovalyuk[a,c], Simone Ferrari[a,d], Oliver Kahl[a], Christoph Nebel[e], Gregory N. Goltsman[c,f], Alexander Korneev[c,g], Wolfram H. P. Pernice[a,d]

[a]Karlsruhe Institute of Technology (KIT), Institute of Nanotechnology (INT), Hermann-von-Helmholtz-Platz 1, 76344 Eggenstein- Leopoldshafen, Germany;
[b]Karlsruhe Institute of Technology (KIT), Institute of Theoretical Solid State Physics (TFP), Wolfgang-Gaede-Str. 1, 76131 Karlsruhe, Germany;
[c]Moscow State Pedagogical University, Department of Physics, Moscow 119992, Russia;
[d]University of Muenster, Institute of Physics, Wilhelm-Klemm-Str. 10, 48149 Muenster, Germany;
[e]Fraunhofer Institute for Applied Solid State Physics (IAF), Tullastr. 72, 79108 Freiburg, Germany;
[f]National Research University Higher School of Economics, 20 Myasnitskaya Ulitsa, Moscow 101000, Russia;
[g]Moscow Institute of Physics and Technology (State University), Moscow 141700, Russia

## ABSTRACT

We report on the design, fabrication and measurement of travelling-wave superconducting nanowire single-photon detectors (SNSPDs) integrated with polycrystalline diamond photonic circuits. We analyze their performance both in the near-infrared wavelength regime around 1600 nm and at 765 nm. Near-IR detection is important for compatibility with the telecommunication infrastructure, while operation in the visible wavelength range is relevant for compatibility with the emission line of silicon vacancy centers in diamond which can be used as efficient single-photon sources. Our detectors feature high critical currents (up to 31 µA) and high performance in terms of efficiency (up to 74% at 765 nm), noise-equivalent power (down to $4.4\times10^{-19}$ W/Hz$^{1/2}$ at 765 nm) and timing jitter (down to 23 ps).

**Keywords:** Single-Photon Detector, Superconducting Detector, SNSPD, Diamond Photonics, Integrated Optics.

## 1. INTRODUCTION

Waveguide integrated superconducting nanowire single-photon detectors (SNSPDs) have attracted considerable attention in the last years[1,2], because they combine high detection efficiency for visible and infrared photons, low dark count rates and low timing jitter[3,4], with the advantages of photonic integrated circuits. Due to their scalable fabrication waveguide integrated single-photon detectors are also emerging as important building blocks for future on-chip quantum optical circuits. For realizing such non-classical circuits the detectors need to be directly structured on top of a material which allows for low loss waveguiding and which would ideally also provide single-photon sources. Diamond is an especially promising material in this respect due to its large bandgap and relatively high refractive index combined with a variety of color centers[5], such as the nitrogen vacancy (NV) and the silicon vacancy (SiV) centers, which are promising single-photon sources. This motivation triggered intense research in recent years on fabricating integrated optical devices out of diamond and also on integrating color centers in diamond optical cavities[6,7], making the scalable integration of single-photon detectors on diamond a remaining challenge.

While waveguide integrated SNSPDs with efficiencies approaching unity and timing jitter below 50 ps have been shown on commercially available materials such as silicon[2] and silicon nitride[8], the first implementations of SNSPDs on diamond[9,10] showed modest detection efficiencies around 50% at 1550 nm, but timing jitters larger than 160 ps, limited by the low critical current below 5 µA. Here we report on the design and experimental demonstration of waveguide integrated superconducting nanowire single-photon detectors (SNSPD) integrated with polycrystalline diamond photonic circuits[11] with on-chip detection efficiencies up to $73.6 \pm 11.0\%$ at 765 nm and a timing jitter down to 23 ps.

*Patrik.Rath@kit.edu

## 2. DEVICE LAYOUT AND FABRICATION

The design of our waveguide integrated superconducting nanowire single-photon detectors (SNSPDs), as shown in Fig. 1a, consists of the following photonic components: a grating coupler is used to guide light from a tunable laser source into a polycrystalline diamond waveguide. A Y-splitter acts as a 50:50 beam splitter and, to the right side, half of the light propagates along a waveguide to a second grating coupler which couples the light out of the chip into a second optical fiber, where it is detected with an external photo detector. This transmission measurement enables us to measure the coupling efficiency and hence to calibrate the photon flux at the detector. The left output of the 50:50 splitter is directed towards the SNSPD, which is electrically connected to gold electrode pads, visible at the top edge of Fig. 1a.

The devices are fabricated via a sequence of electron beam lithography and dry etching steps, as described in detail in our earlier work.[9,12] This includes the structuring of nanowires into niobium nitride (NbN) thin films using HSQ negative resist, as shown in the scanning electron microscope image Fig. 1b. The superconducting NbN films of 4 nm thickness are deposited by DC magnetron sputtering in a nitrogen and argon atmosphere at a discharge current of 350 mA, a nitrogen partial pressure of $2\times10^{-4}$ mbar and a substrate temperature of 850°C. The deposited films show a critical temperature $T_C = 11.4$ K and a transition width $\Delta T_C = 0.4$ K. We measure a sheet resistance of 400 Ohm/sq. at room temperature and critical current densities in the range of $3\times10^6$ A/cm$^2$ to $6\times10^6$ A/cm$^2$ at 4.2 K, which is comparable to NbN films grown on silicon or sapphire substrates.[13]

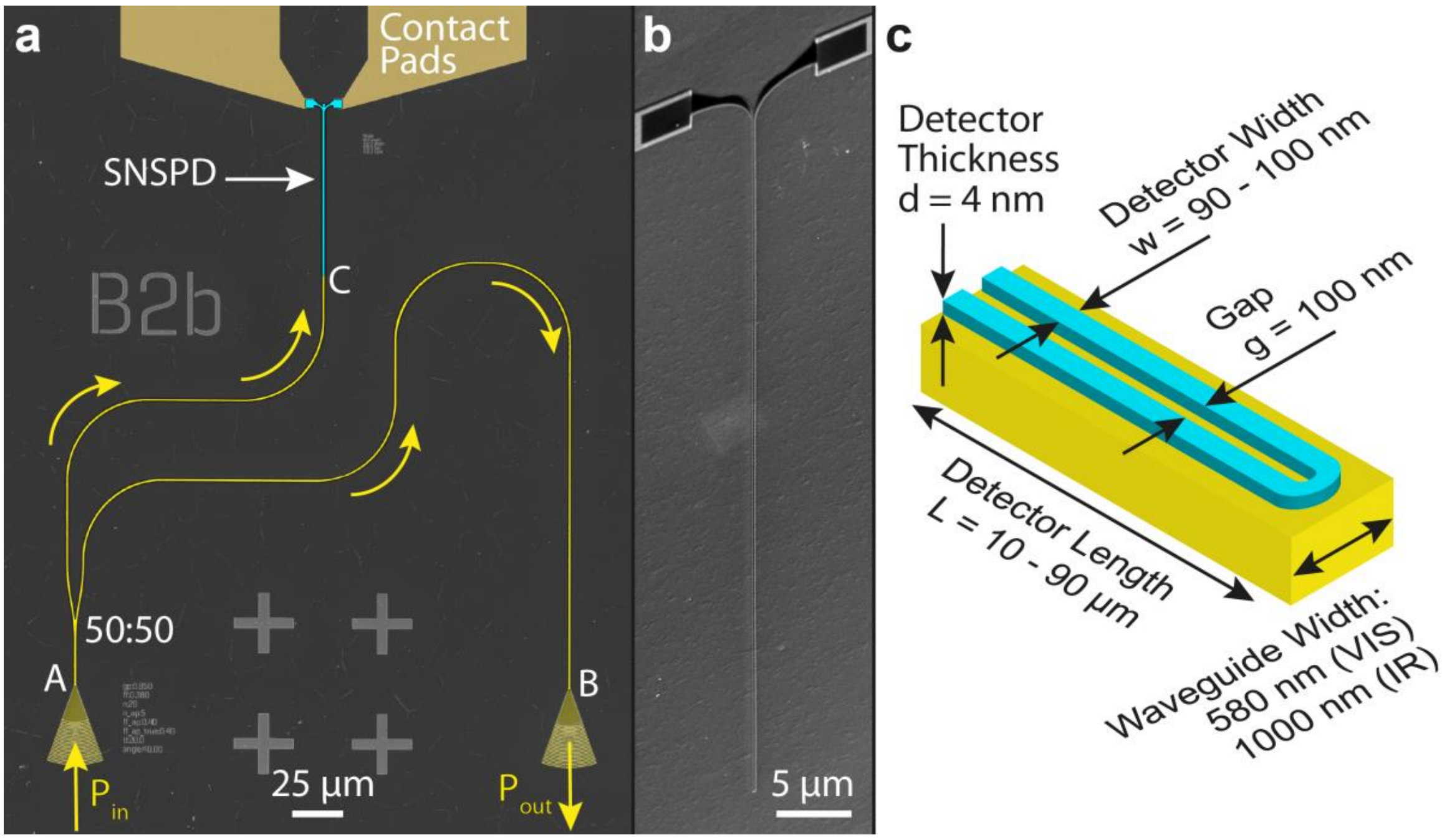

Figure1. Device layout and detector geometry: a) False color scanning electron microscope (SEM) image of a waveguide integrated SNSPD. b) SEM micrograph showing the nanowire geometry of the SNSPD written into negative resist (HSQ 6%) by electron beam lithography during the device fabrication. c) Sketch of the geometric parameters of a waveguide integrated SNSPD, consisting of an NbN nanowire (cyan) on top of a diamond photonic waveguide (yellow).

A sketch of a waveguide integrated SNSPD is shown in Fig. 1c, indicating the main geometric design parameters: the 4 nm thick NbN nanowire (cyan) consists of two parallel stripes of either 90 nm or 100 nm width and a length between 10 µm and 90 µm. These stripes are separated by a 100 nm wide gap and are connected by a 180° circular bend which constitutes the detector's tip. The detector is situated on top of a diamond photonic waveguide (yellow) of 580 nm width, when designed for visible photons (VIS, 765 nm) or 1000 nm width when designed for infrared (IR, 1600 nm). In this configuration the evanescent field of the waveguide mode couples to the NbN nanowire which results in the absorption of photons in the NbN strip. This leads to the breakdown of superconductivity and subsequently enables the detection of the photon.

While this general design has been used in most demonstrations of waveguide integrated SNSPDs to date, one variable in the determination of the efficiency of the detector has to be measured by a separate measurement: the propagation loss of the waveguides. While this can be determined for example via the quality factor of ring resonators or through the cutback method by varying the length of waveguides[11], the propagation loss has to be measured using additional integrated optical circuits on the same chip. By designing the circuits explained above in the right fashion one can avoid this need, which is especially helpful when determining the detection efficiency at several wavelengths, because the propagation loss is wavelength dependent. When designing the length from input to output coupler (section A-B in Fig. 1a) to be twice as long as the distance between input coupler and SNSPD (section A-C), the propagation loss does not contribute to the calculation of the on-chip detection efficiency, as explained in the following.

The on-chip detection efficiency (OCDE) of a detector is obtained by dividing the amount of detection events in a certain time period, called the count rate, by the number of photons impinging on the detector during the same time period, called the photon flux. In our measurements the laser attenuation is adjusted, such that $10^7$ photons per second arrive at the detector. The transmission $T$ through the photonic reference circuit amounts to $T = \frac{P_{out}}{P_{in}} = C^2 \times S \times \exp(-\alpha \times L_{ref})$ where $P_{in}$ is the laser power arriving at the input coupler, $P_{out}$ is the laser power measured after transmission at the output coupler, $C$ is the coupling efficiency of one grating coupler, $S = 0.5$ is the splitting ratio of the Y-Splitter, $\alpha$ is the propagation loss of the waveguide and $L_{ref}$ is the length of the waveguide between the two grating couplers (section A-B in Fig. 1a). The photon flux $\phi$ arriving at the detector is on the other hand given by $\phi = \frac{P_{in}}{E_{photon}} \times C \times S \times \exp(-\alpha \times L_{\det})$, where $E_{photon} = \hbar\,\omega$ is the photon energy and $L_{\det}$ the length of the waveguide between input coupler and SNSPD (section A-B). Because the circuit is designed with $L_{ref} = 2 \times L_{\det}$ this means that $\sqrt{T} = C \times \sqrt{S} \times \exp(-\alpha \times L_{\det})$ and hence $\phi = \frac{P_{in}}{E_{photon}} \times \sqrt{S} \times \sqrt{T} = \frac{P_{in}}{E_{photon}} \times \sqrt{0.5} \times \sqrt{\frac{P_{out}}{P_{in}}}$, which is independent of the propagation loss $\alpha$ and enables us to determine the photon flux simply by measuring the optical input and output power.

We design two sets of devices, for wavelengths of 1600 nm and 765 nm. The respective devices differ only in the geometry of the grating couplers and the waveguide widths, but are equal in all other design parameters. Different wavelengths and waveguide widths lead to different absorption efficiencies because of different modal patterns. Yet because the nanowire geometry is independent of the waveguide geometry we can directly compare the measurement results for the same SNSPD geometry at both wavelengths.

## 3. DETECTION EFFICIENCY AND DARK COUNT RATE - AT 1600NM AND 765NM

We characterize the SNSPDs in a cryogenic setup at a base temperature of 1.8 K, as explained in our earlier work.[2,9] We find that for a nanowire cross-section of 4 nm × 90 nm the NbN nanowires show a maximum critical current of 31 µA, which is about a factor of six larger than in previous work[9] and on par with the best demonstrations of NbN nanowire SNSPDs on other substrates reported in the literature, see comparison by Schuck et. al.[13]

We measure the OCDE in dependence of the bias current for detectors designed for a wavelength of 765 nm (blue circles) and 1600 nm (red triangles). The results for the best out of 16 detectors at each wavelength are shown in Fig. 2. For light with 1600 nm wavelength we find that the OCDE is continuously increasing with bias current, up to the critical current and no efficiency plateau is reached. In this wavelength range we obtain a maximum OCDE of $28.4 \pm 3.4\%$ when biased at 98.8% of the critical current. This efficiency is lower than the 50% shown for comparable detectors in our previous work.[10] We attribute this difference to a larger than designed nanowire cross-section, which reduces the internal quantum efficiency and hence OCDE.

For photons with 765 nm wavelength we find that, for detectors with a high critical current, the efficiency reaches a plateau, as can be clearly seen on the plot in linear scale, as shown in Fig. 2b. For a detector with $L$= 80 µm and $w$ = 90 nm the efficiency plateau reaches a value of $73.6 \pm 11.0\%$ . Reaching a plateau means that a stable operation with low dark count rates is possible because the detectors can be operated at a bias current of about 80% of the critical current without any compromise in terms of efficiency.

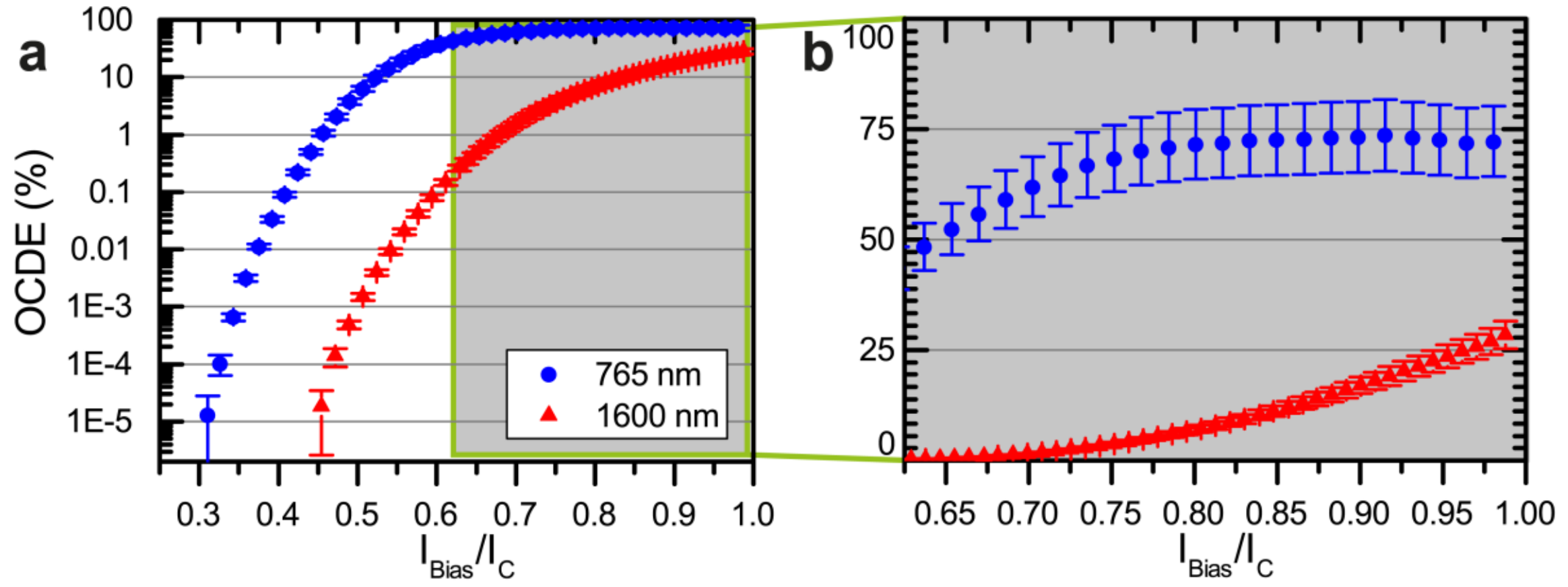

Figure 2. On-chip detection efficiency for the best device at 1600 nm (red triangles) and 765 nm (blue dots). Both nanowires have a width of 90 nm. The detectors lengths of 80 µm (765 nm) and 90 µm (1600 nm) and show critical currents $I_C$ = 31.2 µA (765 nm) and $I_C$ = 28.6 µA (1600 nm), respectively. Measurements were taken at a temperature of 1.8K. a) OCDE in dependence of the bias current $I_B$ relative to the critical current. b) OCDE plotted on a linear scale reveals the plateau in efficiency for photons at 765 nm with a maximum efficiency of $73.6 \pm 11.0\%$ .

We measure the dark count rate (DCR) under the same conditions as the OCDE, only with the fibers between laser source and detectors disconnected. Fig. 3a shows the dependence of the dark count rate on the bias current for the same detectors for which the OCDE was presented in Fig. 2. We find that when reducing the bias current, starting at the critical current, the dark count rate drops exponentially in the high current regime and then slowly decreases below 1 Hz due to remaining stray light and black body radiation in the cryogenic setup.[8]

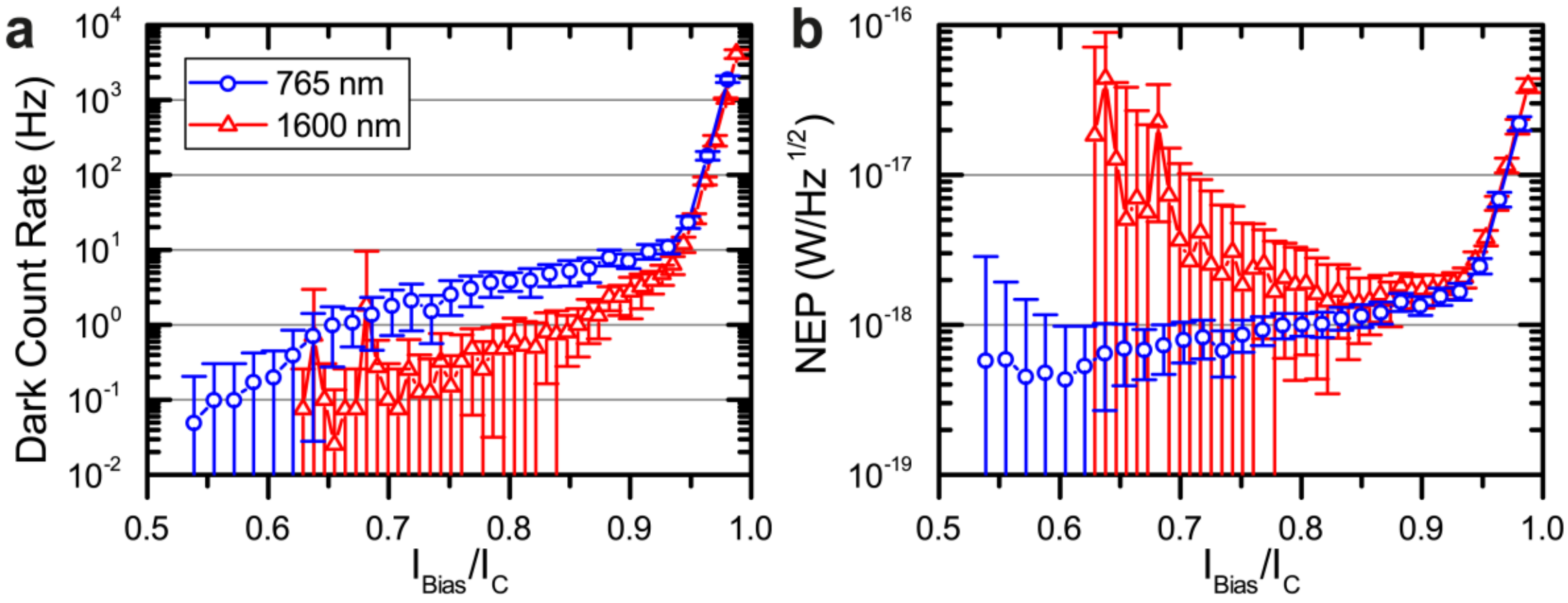

Figure 3. Dark count rate and noise equivalent power: a) Dark count rate as a function of normalized bias current for visible (765 nm) and infrared (1600 nm) photons. b) Noise-equivalent power as a function of normalized bias current for 765 nm (blue) and 1600 nm (red) photons.

From both measurements we extract the corresponding noise-equivalent power (NEP) as shown in Fig. 3b. For NIR photons we obtain minimum NEP of $1.4 \times 10^{-18}$ W/Hz$^{1/2}$ at a normalized bias current of 85%, on par with previous results[9,10], while in the visible wavelength regime the minimum NEP value is reached at a low bias current of 60% and reaches $4.4 \times 10^{-19}$ W/Hz$^{1/2}$. We note that the large error bars towards low bias currents are due to the uncertainty in dark count rates below 1 Hz, due to a data acquisition time of 20 times 2 s per data point in the experiment. The error bars could hence be easily reduced by increasing the measurement time for the dark count rate measurements.

# 4. DECAY TIME AND MAXIMUM COUNT RATE

Next we analyze the detection speed of our devices. We record the electrical pulses resulting from detection events with a 6 GHz digital oscilloscope (Agilent 54855A) in order to infer the decay time. We characterize the same SNSPD, designed for 765 nm with $L$ = 80 µm and $w$ = 90 nm, for which we presented the OCDE and the NEP in the earlier sections. A recorded pulse is shown in blue color in Fig. 4a. An exponential fit (red) reveals a decay time of 2.9 ns, which is limited by the kinetic inductance of the nanowire. We vary the nanowire length and find that the decay time scales linearly with the nanowire length, as expected for the scaling of the kinetic inductance and shown in Fig. 4b. We note that what we call the nanowire length is twice as long as what we call the detector length, because the nanowire consists of two parallel strips connected by 180° turn, as can been seen in Fig. 1c. By reducing the nanowire length to 20 µm a decay time as short as 440 ps is achieved. We note that we do not observe latching[14] of any detector, even for the devices with the shortest decay time of 440 ps.

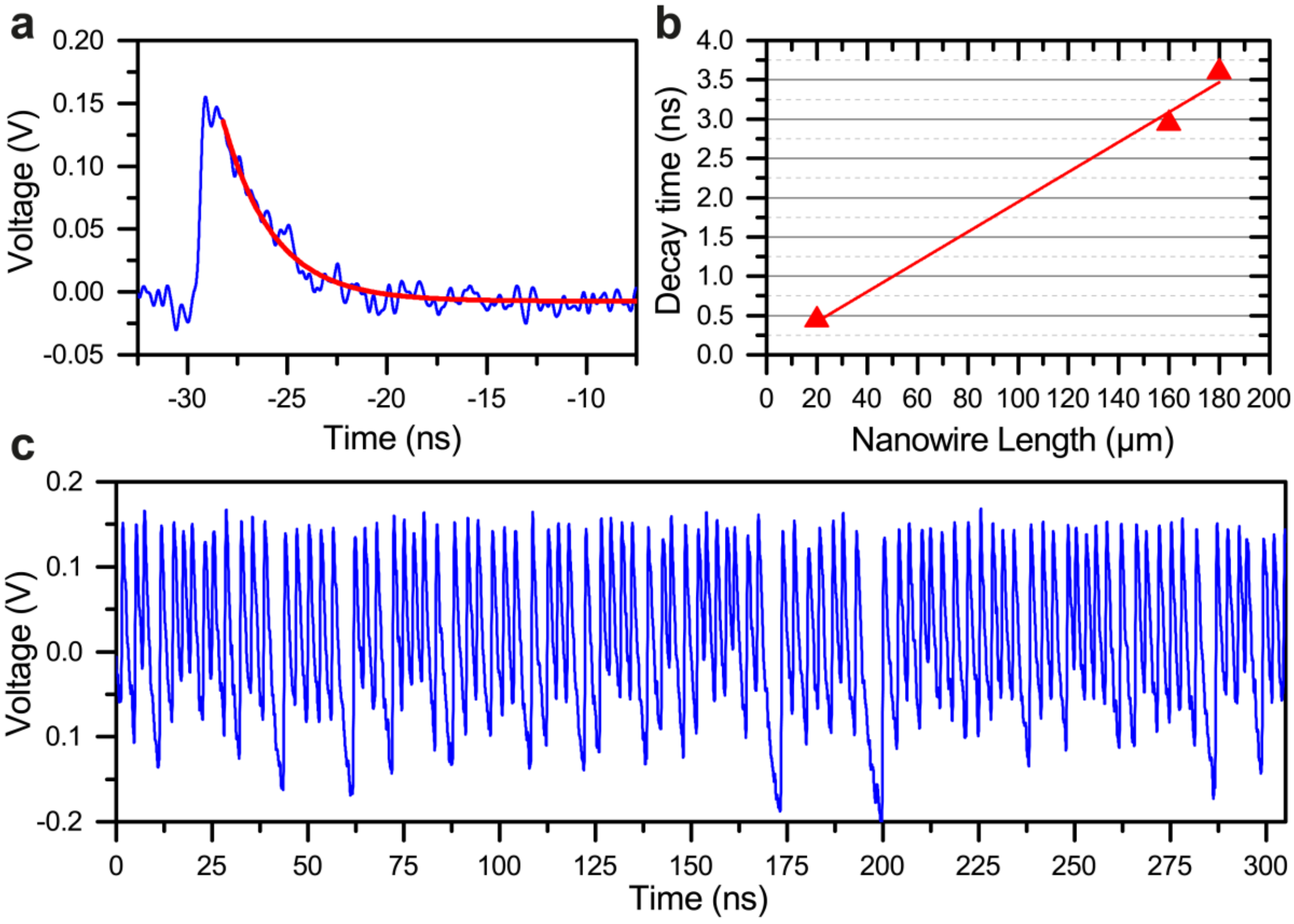

Figure 4. Decay time and maximum count rate of the SNSPDs at 765 nm. a) A measured detection pulse (blue) after 40 dB signal amplification for a detector ($L$ = 80 µm, $w$ = 90 nm), biased at 27 µA. The exponential fit (red) reveals a decay time of 2.9 ns. b) Decay time depending on the nanowire length (twice as long as the detector length) for detectors with 90 nm wire width. The shortest detector yields a decay time of 440 ps. c) One shot recording of the output of the detector biased at 14 µA. The time trace contains 93 pulses within a time interval of 305 ns, corresponding to a frequency of 305 MHz.

We probe the potential to operate the detector at high count rates by stepwise decreasing the attenuation of the attenuators to the input laser in order to increase the laser power of the 765 nm continuous wave (CW) laser which reaches the detector. We test the detector which showed a plateau in OCDE (see Fig. 2) and operate it at a bias current of 14 µA. This particular detector features a decay time of 2.9 ns, which means that a count rate of ~300 MHz should be achievable. We decrease the attenuation and record the count rate depending on the attenuator settings. We find that the count rate steadily increases until it saturates at ~200 MHz, limited by the maximum count rate of the employed frequency counter (Agilent 53132A). Figure 4c shows a one-shot time trace of the detector signal, recorded with the 6 GHz digital oscilloscope, for which the frequency counter yielded a value of 180 MHz. The time trace contains 93 pulses within a time interval of 305 ns, corresponding to a frequency of 305 MHz, which matches with the expected achievable count rate. We note that when operated at the described conditions the detector shows no loss of superconductivity due to latching or indications for problems in free running operation. These measurements illustrate that such detectors provide a large dynamic range and can be operated over at least 8 orders of magnitude in count rate, from their dark count rate of ~1 Hz up to at least $2\times10^8$ Hz. Therefore such SNSPDs can not only be employed for applications where low count rates are expected and hence low dark count rates are needed, but also at high count rates up to hundreds of MHz. In the future this characterization could be improved using a frequency counter allowing higher count rates and by replacing the CW laser with a single-photon source with variable repetition rate >500 MHz. Such an improved system would allow exploring the limits of such detectors at their highest count rates.

## 5. TIMING JITTER

Finally, we characterize the timing performance of the detectors using a pulsed laser source (PriTel FFL- 40M) at 1600 nm wavelength and with 40 MHz repetition rate. We note that the jitter for detectors at 765 nm could not be determined due to the lack of an appropriate pulsed laser source at this wavelength, but we expect the timing jitter to be the same at 765 nm and 1600 nm. The intensity of the laser is adjusted such that less than one photon per pulse on average arrives at the detector, avoiding possible influences of multi-photon detection events.[15] The timing jitter is then determined using a fast oscilloscope in histogram mode, providing a histogram of the time difference between the reference signal of the laser and the detection event of the SNSPD as shown in Fig. 5a. We determine the timing jitter of a typical SNSPD in dependence of the applied bias currents, as shown in Fig. 5b. The height of the output pulse of the SNSPD scales linearly with the bias current and for this measurement we adjust the trigger level to remain at 50% of the average signal height for each bias current.

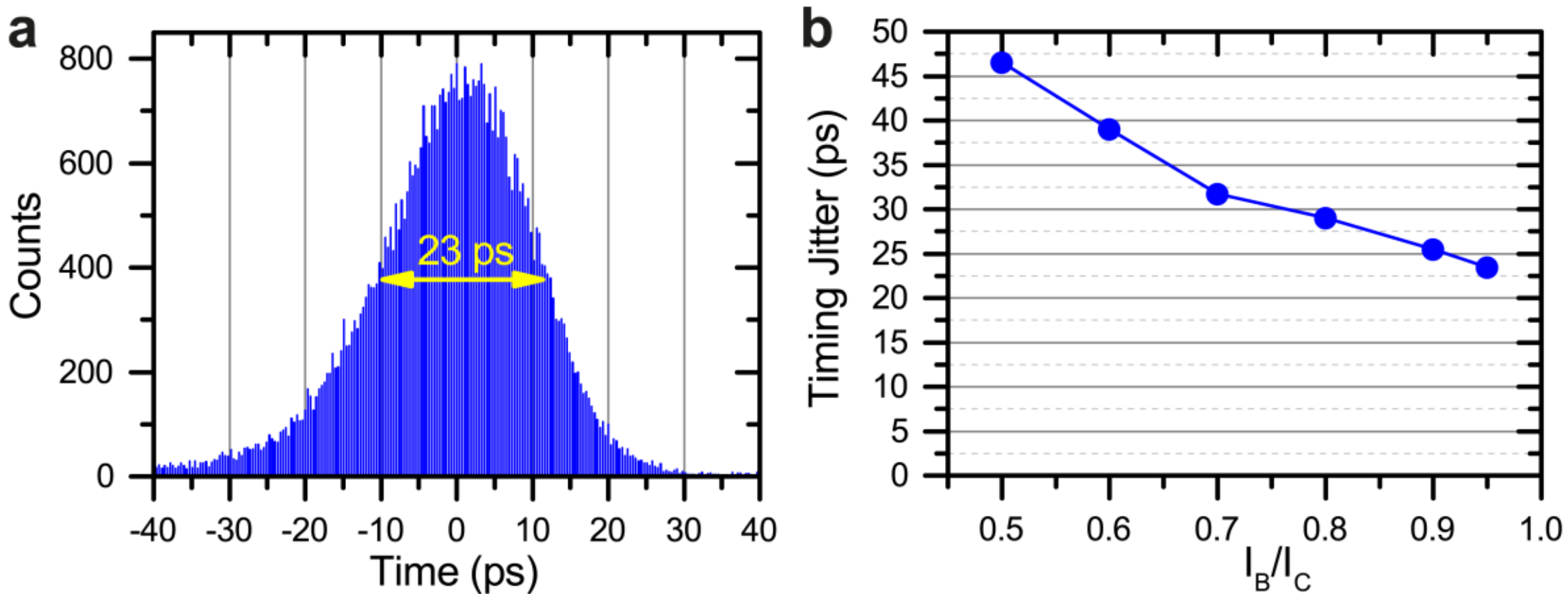

Figure 5. Timing jitter of a SNSPD ($L$= 90 µm, $w$= 90 nm) with a critical current of 28.6 µA. a) Histogram of the arrival times of SNSPD counts relative to a reference signal of the pulsed laser source. The histogram consists of 53,000 total counts at a bin size of 0.36 ps. The SNSPD is biased at a current of 0.95 $I_C$. The Gaussian fit reveals a FWHM value of 23 ps. b) The timing jitter in dependence of the applied bias current $I_B$ relative to the critical current $I_C$.

When increasing the bias current from 50% of the critical current to 95% of the critical current, the jitter continuously decreases from 47 ps to 23 ps. The observed bias dependence of the jitter is similar to those observed for NbN and MoSi SNSPDs[16,17] and can be attributed to the improvement of signal-to-noise ratio with increasing bias current. The jitter for waveguide integrated SNSPDs on diamond was previously limited by a low signal-to-noise ratio, due to small critical currents of the NbN nanowires.[10] The timing jitter of 23 ps presented in this work is a factor of 7 smaller than in our previous results and is now on par with SNSPDs on commercial substrates, such as silicon and silicon nitride.[2,8]

## 6. DETECTOR OPERATION POINT AND ADVANCED CIRCUITS

The best value for each detector property (i.e. OCDE, DCR, and timing jitter) is typically not achieved at the same operation point. Summarizing the experimental results for one SNSPD ($w$ = 90 nm) at one operation point ($T$ = 1.8 K, $I_B$ = 0.93 $I_C$), as provided in table 1, yields the following: the detector has a decay time of 3.6 ns and is able to detect photons of 1600 nm wavelength with an on-chip detection efficiency of 20% at a dark count rate of 5 Hz, which corresponds to a noise-equivalent power $1.9\times10^{-18}$ W/Hz$^{1/2}$. At this operation point the SNSPD has a timing jitter of only 25 ps. At 765 nm wavelength a comparable detector at these operation conditions yields an OCDE of 73% at a dark count rate of 11 Hz, which corresponds to a noise-equivalent power $1.7\times10^{-18}$ W/Hz$^{1/2}$.

Table 1. Performance of waveguide integrated SNSPDs on polycrystalline diamond at a realistic operation point which is maintained for both wavelengths. The working point is a compromise between low DCR, high OCDE and low timing jitter.

| Photon Wavelength | Detector Length | Detector Width | Critical Current $I_C$ | Cryostat Temperature | Bias Current ($I_B$ / $I_C$) | On-Chip Detection Efficiency | Dark Count Rate | Noise-Equivalent Power | Timing Jitter |
|---|---|---|---|---|---|---|---|---|---|
| 1600 nm | 90 µm | 90 nm | 28.6 µA | 1.8 K | 93% | 20% | 5 Hz | $1.9\times10^{-18}$ W/Hz$^{1/2}$ | 25 ps |
| 765 nm | 80 µm | 90 nm | 31.2 µA | 1.8 K | 93% | 73% | 11 Hz | $1.7\times10^{-18}$ W/Hz$^{1/2}$ | -- |

The presented detectors can be directly combined with other optimized integrated optical components developed for diamond integrated optics, such as on-chip interferometers, optical cavities and optomechanical components[18–20] in order to assemble a range of functional photonic circuits on one chip, as shown in Fig. 6a. This will allow for realizing complex single-photon circuits as required for future on-chip quantum optical systems. A prototypical device, shown in Fig. 6b features two SNSPDs on waveguides at the output waveguides of a 50:50 beam splitter, which enables on-chip correlation measurements with fast and efficient readout and small footprint. Such devices are the first step towards combining on-chip quantum interference and single-photon detection in one photonic circuit.[21]

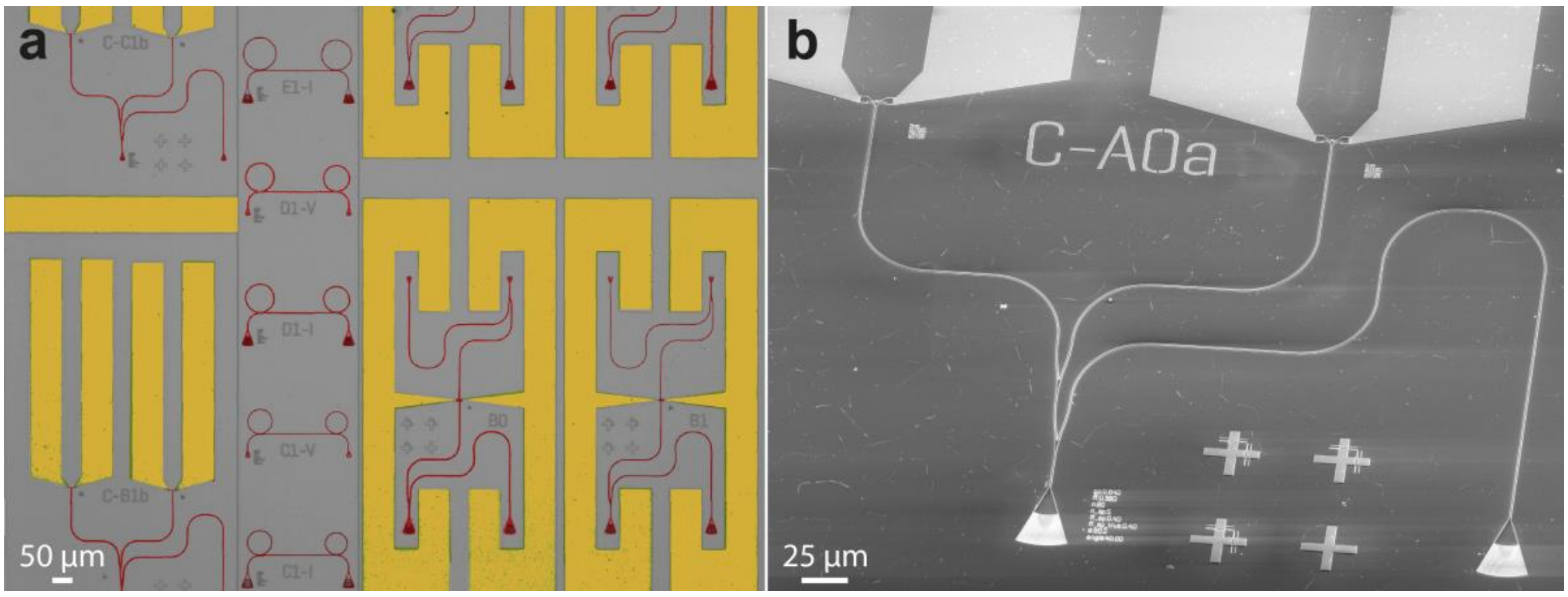

Figure 6. Integration of SNSPDs into functional photonic circuits. a) Colorized microscope image of a photonic chip featuring SNSPDs integrated into various device layouts. b) SEM micrograph of a diamond photonic circuit featuring two travelling-wave single-photon detectors for on-chip correlation measurements.

## 7. CONCLUSIONS AND OUTLOOK

In this manuscript we reported on the design and experimental demonstration of efficient and fast waveguide integrated superconducting nanowire single-photon detectors (SNSPD) integrated with polycrystalline diamond photonic circuits. Our detectors feature high critical current (up to 31 µA) and high performance in terms of efficiency (up to 74% at 765 nm), noise-equivalent-power (down to $4.4\times10^{-19}$ W/Hz$^{1/2}$ at 765 nm) and timing jitter (down to 23 ps). This demonstration of traveling-wave SNSPDs on diamond waveguides is an important step towards the full integration of single-photon sources, photonic circuitry and single-photon detectors on a single chip.

## ACKNOWLEDGEMENTS

Patrik Rath acknowledges financial support by the Deutsche Telekom Stiftung. Wolfram Pernice acknowledges support from the DFG (Grants Nos. PE 1832/1-1 & PE 1832/2-1) and the Helmholtz Society (Grant No. HIRG-0005). Vadim Kovaluk, Gregory N. Goltsman and Alexander Korneev acknowledge financial support from RFBR grant 15-52-10044 and State Contract No. 14.B25.31.0007. The PhD education of Patrik Rath and Oliver Kahl is embedded in the Karlsruhe School of Optics & Photonics (KSOP). We also acknowledge support by the DFG and the State of Baden-Wuerttemberg through the DFG-Center for Functional Nanostructures (CFN). We thank S. Kuehn for help with device fabrication.